\documentstyle[11pt,newpasp,twoside,psfig]{article}
\def\psr{PSR J0218+4232\ }
\def\bll{3C66A\ }
\font\bieleven=cmbxti11
\begin{document}

\title{Likely detection of pulsed high-energy $\gamma$-rays from millisecond
pulsar \psr}
%%% 1 %%%
\author{L. Kuiper, W. Hermsen}
\affil{SRON-Utrecht, Sorbonnelaan 2, 3584 CA, Utrecht, The Netherlands}
%%% 2 %%%
\author{F. Verbunt}
\affil{Astronomical Institute Utrecht, 3508 TA, Utrecht, The 
Netherlands}
%%% 3 %%%
\author{A. Lyne, I. Stairs}
\affil{University of Manchester, Jodrell Bank, Macclesfield SK11 9DL, UK}
%%% 4 %%%
\author{D.J. Thompson}
\affil{Code 661, LHEA, NASA GSFC, Greenbelt, MD 20771, USA}
%%% 5 %%%
\author{G. Cusumano}
\affil{IFCAI CNR, Via U.La Malfa 153, I-90146, Palermo, Italy}

\begin{abstract}
We report on the likely detection of pulsed high-energy $\gamma$-rays from 
the binary millisecond pulsar \psr in 100-1000 MeV data from CGRO EGRET. 
Imaging analysis demonstrates that the highly significant $\gamma$-ray source 
2EG J0220+4228 ($\sim 10\sigma$) is for energies $> 100$ MeV positionally 
consistent with both \psr and the BL Lac \bll. However, above 1 GeV 3C66A is
the evident counterpart, whereas between 100 and 300 MeV \psr is the most likely one. 
Timing analysis using one ephemeris valid for all EGRET observations yields in
the 100-1000 MeV range a double-pulse profile at a $\sim 3.5\sigma$ significance level.  
The phase separation is similar to the component separation of $\sim 0.47$ observed at 
X-rays. A comparison of the $\gamma$-ray profile with the 610 MHz radio profile in 
absolute phase shows that the two $\gamma$- ray pulses coincide with two of the three 
emission features in the complex radio profile.
\end{abstract}

\section{Introduction}
\psr was discovered by Navarro et al. (1995) as a 2.3 ms radio-pulsar in a two day orbit 
around a low mass ($\sim$ 0.2 M$_\odot$) white dwarf companion. A striking feature was that 
the radio-profile appeared complex and very broad.

Targeted observations at soft X-rays (0.1-2.4 keV) with the ROSAT HRI instrument 
revealed also the pulsed nature in the soft X-ray window: a double peaked lightcurve with a 
main emission feature phase separated by $\sim 0.47$ from a second less prominent pulse 
(Kuiper et al. 1998). 

In a recent observation at harder X-rays (1.6-10 keV) with the BSAX MECS instruments the double
peaked nature of the X-ray profile was confirmed (Mineo et al. 2000; Nicastro, these proceedings). 
Spectral analysis showed that the pulsed emission has a very hard spectrum with 
a power-law photon-index of $\sim -0.6$. 

At high-energy $\gamma$-rays the positional coincidence of 2EG J0220+4228 (Thompson et al. 1995)
with \psr was noticed by Verbunt et al. (1996). These authors found also indications for pulsed
emission at energies above 100 MeV. In this work all available EGRET observations of \psr between 
April 1991 and November 1998 with off-axis angles $<30\deg$ have been used to obtain maximum statistics. 
In the timing analysis we used one single very accurate ephemeris (rms error $85\mu s$) having a 
validity interval of about 5 years.    

\section{Imaging analysis}

\begin{figure}[t]
  \centerline{
              \parbox[b]{0.5\columnwidth}
              {\hfill \psfig{figure=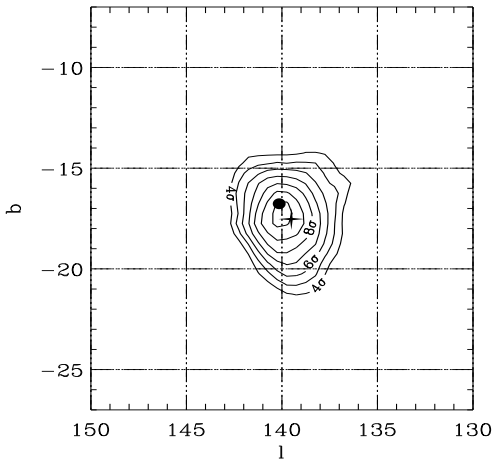,width=7.7cm,height=6.5cm}}
              \parbox[b]{0.495\columnwidth}
              {\caption[]{MLR image in galactic coordinates for energies in excess of 
              100 MeV of the sky 
              region containing 2EG J0220+4228, combining data from 5 separate observations.
              A detection significance of $\ga 10 \sigma$ is reached. The contours 
              start at $4\sigma$ in steps of $1\sigma$ for 1 degree of freedom. 
              \psr is indicated by a star symbol and \bll by a bullet.\vspace{0.55cm}}
              }
             }
   \vspace{-0.5cm}
\end{figure}

We have combined data from CGRO viewing periods 15, 211, 325, 427 and 728.7/9 and binned
the measured $\gamma$-ray arrival directions in a galactic $0 \fdg5 \times 0\fdg 5$ grid
after applying ``standard" EGRET event selections. The measured distribution is compared with an
expected model distribution, composed of galactic and extra-galactic diffuse model components and
established high-energy $\gamma$-ray sources within a $30\fdg 0$ radius around \psr,
by applying a Maximum Likelihood Ratio (MLR) test for the presence of a source at each grid 
position (for more details see Kuiper et al. 2000).

The MLR-map for energies $> 100$ MeV is shown in Fig. 1 with superimposed the positions of 2 
candidate counterparts, \psr and 3C66A. The detection significance of the $\gamma$-ray source 
reaches a $\ga 10\sigma$ level. The number of counts ($> 100$ MeV) assigned to this excess 
is $\sim 230$. We also produced MLR-maps in the broad ``standard'' EGRET differential energy 
windows: 100-300 MeV, 300-1000 MeV and 1-10 GeV. 
The resulting location confidence contours of the $\gamma$-ray source are shown in Fig. 2 for 
all 3 energy windows. 

\begin{figure}
  \centerline{
              \parbox[b]{0.5\columnwidth}
              {\hfill \psfig{figure=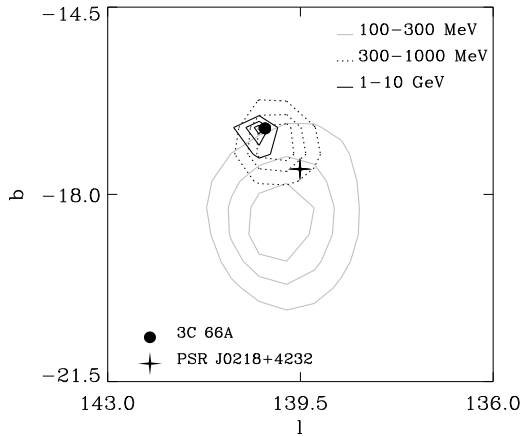,width=7.2cm,height=7cm}}
              \parbox[b]{0.495\columnwidth}
              {\caption{$1,2$ and $3\sigma$ location confidence contours of $\gamma$-ray source
              2EG J0220+4228 in three different broad energy intervals.  Between 100-300 MeV \bll
              is located outside the $3\sigma$ contour, whereas between 1-10 GeV this is the case 
              for \psr.\vspace{1.5cm}}
              }
             } 
  \vspace{-0.5cm} 
\end{figure}

This figure shows that \bll is the evident counterpart for the 1-10 GeV window (consistent 
with the third EGRET catalogue result; Hartman et al. 1999), whereas \psr is the most likely
counterpart for the 100-300 MeV window.  Between 300 and 1000 MeV both sources contribute to
the excess.

\section{Timing analysis}

For the timing analysis we have selected events in a circular aperture around the \psr
position with an energy dependent extraction radius. This radius has been determined a priori
from a signal-to-noise optimization study taking into account the energy dependent point source 
distribution and the best fit total sky background model as derived in the imaging analysis. 
%We also abandonned the requirement of measuring more than 6.5 MeV in either
%of the 2 TASC PHA's as this turned out to be a better selection criterion in a timing analysis of 
%the Crab pulsar. 

We folded the barycentric arrival times of 100-1000 MeV events with the pulsar timing parameters
from one single ephemeris taking into account the binary nature of the system.
We obtained a $3.5\sigma$ signal in a $Z_4^2$-test and the lightcurve showed one prominent
pulse between phases 0.6 and 0.7 following a broader less prominent pulse between phases 0.1 and 
0.4 (see Fig. 3b). In addition, a pulse phase resolved imaging analysis (see Kuiper et al. 2000) shows 
that the 100-300 MeV spatial signal is concentrated in the 2 pulses.

A comparison with the X-ray BSAX MECS and ROSAT HRI lightcurves shows that the phase separation of 
the pulses in the $\gamma$-ray lightcurve is similar to the separation of $\sim 0.47$ found at X-rays.

Finally, we can compare the 100-1000 MeV lightcurve in absolute phase with the 610 MHz radio profile 
(Fig. 3a) and find that the 2 $\gamma$-pulses coincide with 2 of the 3 radio-pulses within the CGRO 
absolute timing uncertainty of $\la 100\mu s$.

\section{Summary}

This study shows that we obtained good circumstantial evidence for the first detection of
high-energy $\gamma$-rays from a millisecond pulsar, \psr:

\begin{itemize}

\item[$\bullet$] A double-peaked lightcurve in the 100-1000 MeV energy interval with a 
$\sim 3.5\sigma$ modulation significance.

\item[$\bullet$] The phase separation of the 2 $\gamma$-pulses is similar to that at
hard X-rays; a comparison in absolute time with the 610 MHz radio-profile shows alignment of
the $\gamma$ pulses with 2 of the 3 radio pulses.

\item[$\bullet$] Between 100 and 300 MeV the EGRET source position is consistent with \psr
with the signal concentrated in 2 pulses.  
\end{itemize}

The full analysis and implications of our findings will be presented in detail in Kuiper et al. 2000.

\begin{figure}
  \centerline{
              \parbox[b]{0.5\columnwidth}
              {\hfill \psfig{figure=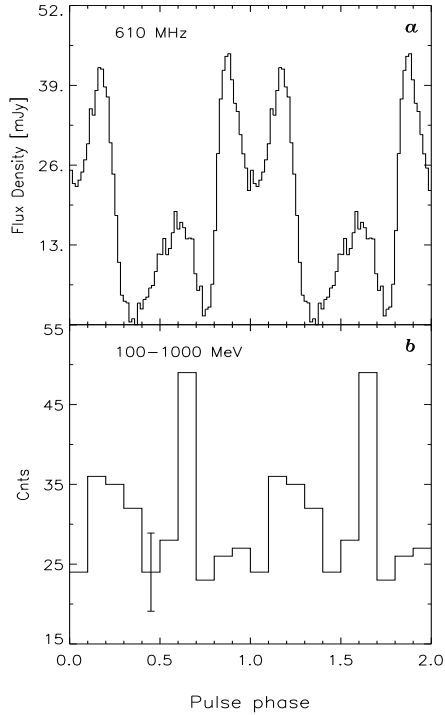,width=6cm,height=10.6cm}}
              \parbox[b]{0.495\columnwidth}
              {\caption{Comparison in absolute time of the radio 610 MHz profile ({\bieleven a}; 
               Stairs et al. 1999) and the 100-1000 MeV  EGRET lightcurve ({\bieleven b}) of 
               \psr. A typical error is indicated in the lower panel. Notice the (near) 
               alignment of the 2 high-energy pulses with 2 of the 3 radio-pulses.\vspace{3.5cm}}
              }
             }
  \vspace{-0.35cm}  
\end{figure}

\end{document}